\documentclass[DIV=calc,paper=a4,fontsize=11pt,twocolumn]{scrartcl} 

\usepackage[english]{babel}
\usepackage[protrusion=true,expansion=true]{microtype}
\usepackage{amsmath,amsfonts,amsthm}
\usepackage{amssymb,stmaryrd,relsize}
\usepackage[final]{graphicx}
\usepackage{xcolor}
\usepackage[normal,small,hypcap,up,labelfont=bf,textfont=it]{caption}
\usepackage{subfig}
\usepackage{booktabs}
\usepackage{fix-cm}
\usepackage{dsfont}
\usepackage{bbm}
\usepackage{pstricks}
\usepackage{cite}
\usepackage[utf8]{inputenc}
\usepackage[perpage,symbol*]{footmisc}
\usepackage[varg]{txfonts}
\usepackage{balance}
\usepackage{fancyhdr}
\PassOptionsToPackage{hyphens}{url}\usepackage[pdfencoding=auto,psdextra]{hyperref}
\usepackage{bookmark}
\usepackage{verbatim}
\usepackage{fontenc}
\usepackage{cuted}

\DeclareCaptionFont{mycolor}{\color[HTML]{000000}}
\def\la{{\langle}}
\def\u{\hat U}
\def\A{\mathcal A}

\def\B{\hat{B}}

\def\R{\text{Re}}
\def\Im{\text{Im}}

\def\lm{\lambda}

\def\q{\quad}

\def\n{\\ \nonumber}
\def\ra{{\rangle}}

\def\l{b}

\usepackage{sectsty}													
\allsectionsfont{
\color[HTML]{31ADF3}\usefont{OT1}{phv}{b}{n}
}

\sectionfont{
\color[HTML]{31ADF3}\usefont{OT1}{phv}{b}{n}
}

\usepackage{fancyhdr}												
\usepackage{lettrine}
\newcommand{\initial}[1]{%
\lettrine[lines=3,lhang=0.3,nindent=0em]{
\color[HTML]{31ADF3}
{\textsf{#1}}}{}}

\usepackage{titling}															

\newcommand{\HorRule}{\color[HTML]{31ADF3}
\rule{\linewidth}{1pt}%
}

\pretitle{\vspace{-30pt} \begin{flushleft} \HorRule
\fontsize{34}{34} \usefont{OT1}{phv}{b}{n} \color[HTML]{31ADF3} \selectfont
}
\title{On Weak Values and Feynman's Blind Alley}					
\posttitle{\par\end{flushleft}\vskip 0.5em}

\preauthor{\begin{flushleft}\large \lineskip 0.5em \usefont{OT1}{phv}{b}{sl} \color[HTML]{31ADF3}}
\author{Dmitri Sokolovski\\[8pt]}											
\postauthor{\footnotesize \usefont{OT1}{phv}{m}{sl} \color[HTML]{000000}
Department of Physical Chemistry and EHU Quantum Center, University of the Basque Country, Leioa, Spain.\\
IKERBASQUE, Basque Foundation for Science, Bilbao, Spain. E-mail: \href{mailto:dgsokol15@gmail.com}{dgsokol15@gmail.com}\\[10pt]		
\scriptsize\usefont{OT1}{phv}{m}{n} \color[HTML]{31ADF3}{\textbf{Editors: \emph{Alexandre Matzkin} \& \emph{Danko D. Georgiev}} }\\[5pt]
\color[HTML]{000000}{Article history: Submitted on September 18, 2023;  Accepted on November 16, 2023; Published on November 19, 2023.}
\par\end{flushleft}\HorRule}

\date{}																				

\begin{document}
\maketitle
\thispagestyle{fancy} 			
\initial{F}\textbf{eynman famously recommended accepting the basic principles of quantum mechanics without trying to guess the machinery behind the law. One of the corollaries of the Uncertainty Principle is that the knowledge of probability amplitudes does not allow one to make meaningful statements about the past of an unobserved quantum system. A particular type of reasoning, based on weak values, appears to do just that. Has Feynman been proven wrong by the more recent developments? Most likely not.\\ Quanta 2023; 12: 180--189.}

\begin{figure}[b!]
\rule{245 pt}{0.5 pt}\\[3pt]
\raisebox{-0.2\height}{\includegraphics[width=5mm]{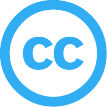}}\raisebox{-0.2\height}{\includegraphics[width=5mm]{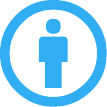}}
\footnotesize{This is an open access article distributed under the terms of the Creative Commons Attribution License \href{http://creativecommons.org/licenses/by/3.0/}{CC-BY-3.0}, which permits unrestricted use, distribution, and reproduction in any medium, provided the original author and source are credited.}
\end{figure}

\section{Introduction}
\label{sec:1}

Ten years ago I published a paper questioning what is known as the \emph{weak measurements} technique \cite{DSqua}.
In this invited contribution, I have the opportunity to do the same again. The subject 
has since progressed (see, e.g., Refs. \cite{WV1,WV2}), certain mistakes were made \cite{WVcrap}, and the opinions remain polarized.
There are those who dismiss ideas like \emph{separating electron from its charge} \cite{WVcat} as pure nonsense,
and those who consider them as significant achievements. Both the enthusiasts of the approach and its detractors (including the present author)
have had their share of good time discussing the matter. The required mathematics is elementary, and the issue appears to go to the very heart of quantum theory.
The sceptic's position is as follows. In quantum mechanics there are two kinds of quantities, complex valued amplitudes, 
and probabilities, quadratic in the former. Measured \emph{weak values} are essentially amplitudes, and amplitudes are not good 
for trying to determine a quantum system's past. This conclusion can be deduced, for example, from the textbook discussion of the basic quantum
principles given in \cite{FeynL}. But what happens if the warning is ignored? We we will discuss this in what follows.
We hope the reader would not mind the use of strong words such as \emph{orthodoxy} and \emph{dogma}, wherever they are needed to emphasise the importance of the concept in question. 
A similar excuse is offered for citing sources from beyond the ambit of quantum physics. 
Some of the material in the Appendices is well known, and is included to facilitate the narrative. 

\section{The reward of the orthodox}
\label{sec:2}

Man, said G.~K.~Chesterton \cite{CHEST}, can be defined as an animal that makes dogmas. 
Unlike Chesterton, Feynman had little sympathy for catholic orthodoxy \cite{FeynP}, 
but appeared no less stringent while laying out the principles of quantum physics \cite{FeynL}.

Without going into the technical details, these rules say that there are many elementary (virtual) scenarios of what may happen to a quantum system. 
To each scenario quantum theory ascribes a complex number known as a probability 
amplitude. If a (real) sequence of observed events is consistent with several virtual scenarios, 
the absolute square of the sum of the corresponding amplitudes gives the probability with which the observed sequence will
appear after many identical trials. The amplitudes are never added for distinguishable \emph{in principle} final conditions. 
This simple recipe captures \emph{the only mystery} of quantum mechanics \cite{FeynL}, the new content which distinguishes 
it from classical physics. The phenomenon of interference is best illustrated on the generic double-slit example, where the system starting from an initial state, $I$, can reach a final state, $F$, via two routes, endowed with amplitudes $\A_1$ and $\A_2$. With no attempt to determine which route has been taken, the probability to have the condition $F$ is $P(F\gets I)=|\A_1+\A_2|^2$ 
and, with such an attempt successfully made, this changes to $P'(F\gets I)=|\A_1|^2+|\A_2|^2$.
This is summarized by the Uncertainty Principle, stating that one cannot know the route taken, and keep the interference pattern [i.e., maintain $P(F\gets I)$ intact]. 

This is, of course, well known, and the question is how important can it really be for understanding quantum physics?
Very important, according to Feynman \cite{FeynC}. The mathematics is so simple that no deeper insight into \emph{how can it be like that} is possible.
One can only admit that nature \emph{does behave like this.} Accepting this as a sort of {dogma}, one \emph{will find [nature] a delightful, entrancing thing} \cite{FeynC}, and, we add, have a happy productive life as a quantum physicist. If this is reward of the orthodox, what is the punishment prepared for a heretic? 

\section{The punishment of the heretic}
\label{sec:3}

Truths turn into dogmas the minute they are disputed \cite{CHEST}. The punishment is both self-inflicted and harsh.
 Whoever wishes to go beyond the Uncertainty Principle, or explain in more detail \emph{the machinery behind the law} will \emph{get down the drain}, and find him/herself, together with other unfortunates, in a \emph{blind alley} \cite{FeynC}. A \emph{heretic} is bound to say things which make little sense and, more precisely, make \emph{wrong predictions} \cite{FeynL}. 

For example, assuming that the system is pre-destined to take one of the two routes, suggests that by plugging one of the slits
one can only have fewer particles arriving at a chosen point on the screen. However, with only one slit open, the probability becomes
$P''(F\gets I)=|A_1|^2$, and since there are no \emph{a priori} restrictions on $\A_1$ and $\A_2$, it can happen that $P''(F\gets I) >|\A_1+\A_2|^2$.
This provides an elementary proof of incompatibility of a local hidden variable theory (see, e.g., \cite{Hidd}) with quantum mechanics \cite{FeynC}.
Neither can one say that the system (a~particle) has split into two in order to travel both paths simultaneously, since no one has ever observed half of an electron emerging from one of the slits \cite{FeynC}.

\section{Things better not said}
\label{sec:4}

There is, however, one difficulty with the above arguments, easily seen by an attentive opponent.
The predictions refer to an unobserved and, therefore, unperturbed system. Yet they are disproven by considering 
a system strongly perturbed by the measurement. Not comparing like with like, leaves some room
for discussing what happens if the interference is left intact. Feynman's orthodoxy can be seen as a variant 
of the Copenhagen interpretation \cite{Cop}, which only gives answers to operationally posed questions, and 
could, therefore, be missing other important things. 

There is one well known theory which allows one to trace the path of the particle in double slit experiment.
In Bohmian mechanics \cite{Bohm} one solves the time dependent Schr\"odinger equation for the wave function $\psi(x,t)$, and 
construct the probability field as per usual, $p(x,t)=|\psi(x,t)|^2$. The probability is conserved, so one can construct non-intersecting flow lines, 
identified as the particle's actual trajectories. The scheme offers some gratification for anyone worried about the lack of description of quantum particle's past. In the double slit case, particle arrives at each point on the screen ($F\to x$) via one slit. However, 
this simply places the novel content of quantum theory elsewhere. 
In order to comply with quantum results, something in the classical description must give in and 
the casualty, in this case, is locality. The particle can be seen as leaving the source with a prescribed instruction which trajectory to follow, yet it remains affected by what happens at the other slit, which is not supposed to visit.
Still, one is able to obtain quantum statistics from a picture where the particle follows a continuous trajectory, endowed with a probability, rather 
than with a probability amplitude. 

In the 1964 Messenger Lectures, Feynman seems to soften his stand by conceding that
\begin{quote}
You can always say it [that the particle goes through either one hole, or the other] -- provided you stop thinking immediately and make no deductions from it. Physicists prefer not to say it, rather than to stop thinking at the moment.
\mbox{\cite[p.~144]{FeynC}}
\end{quote}
It is not clear whether Feynman's remark refers to Bohmian mechanics, not mentioned directly in \cite{FeynC}, 
but it is by no means impossible. 

Here we used the Bohmian example to stress that whenever a new interpretation or extension of quantum theory is proposed, it needs to be measured up against the Uncertainty Principle which, according to \cite{FeynC} \emph{can be used to guess ahead at many of the characteristics of unknown objects}.
 Such a comparison may announce the arrival of a better and deeper theory (very good). Or it can confirm the validity of Feynman's \emph{orthodoxy} (good). It can also expose the researher's position inside the proverbial \emph{blind alley}, reserved for the \emph{heretic} (whatever this may mean, not so good). 
A comparison of this kind is long overdue in the case of \emph{weak measurements}, 
and we will try to make it next (see also \cite{DSadp}).

\section{ The \emph{weak values}}
\label{sec:5}

Leaving the technical details aside (cf.~\hyperref[appa]{Appendix A}) we fast forward to the moment when the experimenter, who coupled an inaccurate 
\emph{weak} pointer to a pre- and post-selected two-level system, has 
succeeded in determining both the real and 
imaginary parts of a complex quantity 
\begin{align}
\label{0}
& \la \B \ra_W =\frac{\sum_{j=1}^2 B_j\A(F\gets \l_j\gets I)}{\sum_{j=1}^2\A(F\gets \l_j\gets I)}, \n
& \A(F \gets \l_j\gets I)\equiv \la F|\u(t_2,t_1)|\l_j\ra \la \l_j|\u(t_1)|I\ra,
\end{align}
called the \emph{weak value of operator $\B=\sum_{j=1}^2 B_j|\l_j \ra\la \l_j|$} \cite{WVtod}. 
What new, if anything, has been learned?

\begin{figure*}[t!]
\includegraphics[width = 153mm]{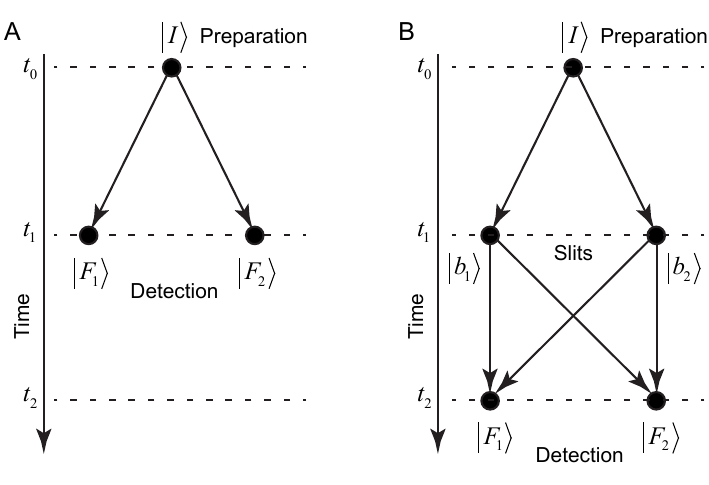}
\centering
\caption{\label{fig_1} (A) Two-step measurement: a two-level system ($N=2$) is first prepared in 
$|I\ra$, and later detected in $|F_i\ra$. Path probabilities 
$P(F_i\gets I)$ always exist [cf. Eq.(\ref{A-2})], and there is no interference to destroy.
(B) Three-step measurement. The probabilities of detection in $|F_i\ra$, $P(F_i\gets I)$ always exists, 
but vary depending on the accuracy (\emph{strength}) of measuring $\B$ at $t=t_1$.
Individual path probabilities $P(F_i\gets \l_j\gets I) =|\A(F_i\gets \l_j\gets I)|^2$ exist
only if an accurate measurement of $\B$ destroys interference between the paths.}
\end{figure*}

It is easy to recognize the setup as a rudimentary double-slit problem, where the states $\l_j$ and $F$ play the role of two \emph{slits}
and and one of the two \emph{points on the screen}, respectively (see Fig.~\ref{fig_1}B).

The expression in Eq.(\ref{0}) is a particular combination of the amplitudes $\A(F\gets \l_j\gets I)$, defined for the four scenarios available to the system. The orthodox view of Section~\ref{sec:2} is that such amplitudes are good only for calculating probabilities, defined as their (or their sum's) absolute squares. Clearly, the complex valued \emph{weak value} (\ref{0}) is neither a probability, nor a conventional average. Could this be a chance to learn something Feynman's orthodoxy has missed? 

We note from the start that it is, however, unlikely. In essence, Feynman argues that one can always know the probability amplitudes, but is still unable to conclude whether the has system passed through one of the slits, or through both. Since the amplitudes are known, one can also know the expressions in the r.h.s. of Eq.(\ref{0}). What could be special about them? 

For one thing they can be measured in a laboratory \cite{WVtod}.
However, Eq.(\ref{0}) does not define a new way of calculating quantum mechanical averages. Rather, one measures the average position (reading) of the pointer in the standard manner, and uses this conventional average to work out, e.g., the real part of $\la \B \ra_W$
(see \hyperref[appa]{Appendix~A}). 
This is neither new, nor particularly unusual. Response of a quantum system to a small perturbation 
 is always expressed in 
terms of the system's amplitudes (see \hyperref[appb]{Appendix~B} for a simple example). 
If the \emph{weak values} do describe \emph{new physics}, as was suggested in \cite{WVtod}, they must also have a truly new physical meaning. 
We will consider their possible interpretations after making sure that we, and the authors of \cite{WVtod}, are indeed talking about the same thing. 

 

\section{A historical note}
\label{sec:6}

The history of \emph{weak values} can be traced back to 1964 when the authors of \cite{ABL} evaluated intermediate mean value of an operator 
$\B=\sum_{j=1}^N B_j |b_j\ra\la b_j|\equiv \sum_{j=1}^N B_j \hat \pi_j$ 
for a system pre- and post-selected in the states $|I\ra$ and $|F\ra$, 
\begin{align}
\label{a0}
\la \B\ra &= \frac{\sum_{j=1}^N B_j |\la F(t_1)| \hat \pi_j|I(t_1)\ra|^2}{\sum_{j=1}^N |\la F(t_1)| \hat \pi_j|I(t_1)\ra|^2}\\
&= \frac{\sum_{j=1}^N B_j|\A(F\gets \l_j\gets I)|^2}{\sum_{j=1}^N|\A(F \gets \l_j\gets I)|^2}, \nonumber
\end{align}
where $|I(t')\ra= \u(t_1)|I\ra$, and $|F(t')\ra= \u^{-1}(t_2-t_1)|F\ra$. The \emph{Aharonov--Bergmann--Lebowitz rule}
[the first equality in (\ref{a0})] is clearly an example of the Feynman's rule for assigning probabilities \cite{FeynL}
[the second equality in (\ref{a0})] in the case where all scenarios can be distinguished. Indeed, it can be obtained as a mean reading 
of an accurate pointer, $\la f(F_i)\ra$ by using Eq.(\ref{A2}). Calculating the same average for a weakly coupled pointer, 
the authors of \cite{WVtod} obtained a somewhat similar expression, 
\begin{align}
\label{a1}
\frac{\la f(F_i) \ra}{\beta} & =  \R\left [\frac{\sum_{j=1}^N B_j \la F(t_1)| \hat \pi_j|I(t_1)\ra}{\sum_{j=1}^N \la F(t_1)| \hat \pi_j|I(t_1)\ra}\right]\n
& =\R\left [\frac{\sum_{j=1}^N B_j \A(F\gets \l_j\gets I)}{\sum_{j=1}^N \A(F\gets \l_j\gets I)}\right] \n
& \equiv \R\left [\la \B\ra_W\right]
\end{align}
where $\beta$ is the coupling strength.
(Noteworthy, a \emph{weak} pointer, $H_{int}=-i\beta\partial_f\B \delta(t-t_1)$, $G(f)$, $\beta \to 0$,
is transformed into inaccurate one by a transformation $f'=f/\beta$, $H'_{int}=-i\partial_f\B \delta(t-t_1)$, 
$G'(f')=G(\beta f')$, whereby $\Delta f'= \delta f/\beta \to \infty$.
 This would yield Eqs.(\ref{a0}) and (\ref{0}) for the \emph{strong} and \emph{weak} regimes, respectively.)
The quantities in the brackets are two equivalent forms of \emph{weak value} $\la \B \ra_W$, introduced earlier in Eq.(\ref{0}) for $N=2$.

\section{Looking for the meaning of the \emph{weak values}}
\label{sec:7}

To an orthodox this search is the most important endeavor of the whole saga.
It is also the main purpose of this paper. 
A pre- and post-selected system, coupled to a pointer, is described by the transition amplitudes 
\mbox{$\A(F_i \gets \l_j \gets I)$} [see Eq.(\ref{A1})], so whatever can be learned about the pointer will always be expressed
in terms of these amplitudes. The \emph{weak value} in Eq.(\ref{0}) is a combination of amplitudes, and 
a kind of amplitude itself. Believing in the Uncertainty Principle, the orthodox also believes that knowing the amplitudes 
(which are always available) can provide no insight into how a particle goes through two slits, or, more generally, 
into a quantum system's past. If the \emph{weak values} are able to shed a new light on this vexed issue, 
Feynman's warnings come to nothing, and the orthodox view of the theory does require a radical overhaul. 
But if they cannot do so, and Feynman was right, the \emph{weak measurements} can lead one into
the by now proverbial \emph{blind alley}. 

As yet there is little clarity as to the physical meaning of a weak value,
 although the real part can be written as a particular conditional expectation value \cite{Matz2}. 
Below we will question
(without prejudice to practical utility of the \emph{weak measurement} technique)
 several propositions which can be found in the literature.
We will test them on the \emph{double slit problem} in Fig.~\ref{fig_1}B,
choosing the operator $\B$ in Eqs.(\ref{0}) and (\ref{a1}) to be a projector onto the first of the two states $|b_j\ra$, 
\begin{equation}
\label{b0}
\B =\hat \pi_1=|\l_1\ra\la \l_1|, \q B_1=1, \q B_2=0.
\end{equation}

\subsection{ A weak value represents the mean value of a variable with interference intact (?)}

References \cite{WVtod}, \cite{WVneg}, come close to suggesting it, when it was argued 
that a \emph{negative kinetic energy} can be attributed
to a particle in a classically forbidden region. 
On the other hand, tunnelling is an interference phenomenon (see, e.g., \cite{DSx}), and the orthodox view implies that the value of kinetic energy, like the slit chosen by the particle, must remain indeterminate.
Indeed, measuring the accurate (\emph{strong}) mean value of any $\B$ is in itself a \emph{which way} problem, 
as is seen from our double slit example. 
After $K\gg 1$ trials, in which the system is always found in $|F\ra$ at $t_2$, one counts the number of times, $N_1$, the system takes the route $F\gets \l_1\gets I$, 
and evaluates a sum $\la \B \ra = [K_1 B_1+(K-K_1)B_2]/K$ which agrees with (\ref{a0})
as $K\to \infty$.
An accurate mean value of $\hat \pi_1$ in (\ref{b0}) is, therefore, the conditional probability of taking the route passing via $|\l_1\ra$ at $t_1$, 
given that the system arrives in $|F\ra $ at $t_2$. 

A \emph{weak} pointer does not distinguish between the two scenarios, and the number of times the route $F\gets \l_1\gets I$ is taken is not only known, 
but, according to the Uncertainty Principle, cannot even be defined in a meaningful way. This given, something must go wrong with treating $\la \hat \pi_1\ra_W$ as a conditional probability, 
and it is easy to see what. The projector's \emph{weak value} 
\begin{equation}
\label{b1}
\la \hat \pi_1 \ra_W =\frac{\A(F\gets \l_1\gets I)}{\sum_{j=1}^2\A(F\gets \l_j\gets I)},
\end{equation}
is an amplitude (renormalised, but still an amplitude) and may take complex values. 
One does not want to say that the system travels a route in $(1+i)K$ out of $K$ cases, since it is not clear what it means.
The real and imaginary parts of $\la \B \ra_W$ are also poor candidates for the role, since they can take either sign, and exceed unity.
The modulus $|\la \B \ra_W|$ is, indeed, positive but may be too large $|\la \B \ra_W|\gg 1$ (see \hyperref[appc]{Appendix~C}).
 In brief, $\la \B \ra_W$ in Eq.(\ref{b1}) cannot be used to determine the relative frequency with which a system, unobserved
 at $t=t_1$, arrives in $|F_1\ra$ via the route $F\gets \l_1\gets I$. The same difficulty occurs with the \emph{weak value}
 of a more general operator, $\la \B \ra_W=\sum_{j=1}^N B_j\la \hat \pi_j\ra_W$.

The orthodox is not surprised. He knows that quantum amplitudes are not suited for making \emph{which way} predictions,
and does not expect a \emph{weak measurement} to yield a further insight into the double slit conundrum.

\subsection {The weak value of a projector is an \emph{occupation number} (?)}

This suggestion was made in \cite{WVh} which analysed
 a \emph{four-slit} problem with path amplitudes $\A_1=0$, and $\A_2=\A_3=-\A_4$, and later in \cite{WVtod} 
dealing with a three-slit problem with $\A_1=-\A_2=\A_3$. 
In both cases it was proposed that there can be negative number of particles, or particle pairs, passing through an arm of an interferometer. 

The idea is also easily tried on the double slit problem. To avoid encountering even more mysterious complex \emph{occupation numbers} one can make 
all the amplitudes in (\ref{C0}) real by putting $\phi=\phi'$. Now choosing $\epsilon=10^{-5}$, and following \cite{WVtod} and \cite{WVh}, one needs
$N_1=10^5$ particles (copies of the system) in the path $F_1\gets \l_1\gets I$, and $N_2=-99999$ particles in the path $F_1\gets \l_2\gets I$. 
An explanation usually consists of describing a new phenomenon in terms of previously defined concepts. 
The above is a hardly an explanation of the double slit phenomenon, since the concept of having a negative number of particles is itself 
undefined. 

 For an orthodox the measured values $\la \hat \pi _1\ra_W=10^5$ and $\la \hat \pi _2\ra_W=-99999$ only reflect the correct relations 
 between the amplitudes $\A(F_1\gets \l_1\gets I)$ and $\A(F_1\gets \l_2\gets I)$, whilst the search for a deeper meaning of these numbers { must} end in a blind alley. And so it does. 
 Explaining detection of the system in its final state as a result of a conspiracy between thousands of copies of the same system may 
 seem a little too extravagant.
Forget, therefore, numerical values. Perhaps the mere fact that a \emph{weak value} does not vanish has a clearer meaning?
 
\subsection {None-zero weak value of a projector indicates presence of the system at the chosen location (?)}

 It does, say the authors of \cite{WVcat}. So much so, that a system and its particular property can part company 
 and go their different ways. One, for example, can \emph{separate [...] internal energy of an atom from the atom itself} \cite{WVcat}.
For the double slit example a similar proposition means
 that whenever neither amplitude vanishes, the particle goes through both slits at once. However, according to \cite{FeynC},
 this directly contradicts the Uncertainty Principle, as well as 
the experimental evidence, since wherever one looks, he finds either entire system, or nothing. 

The orthodox may also add that, as in the case of the Bohmian particle, there is a problem with locality. 
 A final state $|F_i\ra$ can be reached via two paths (arms of the interferometer in an optical realization 
 of the experiment), each endowed with an amplitude $\A(F_i\gets\l_j\gets I)$, $j=1,2$. A property, local to the first arm, can be expected to depend only 
 on one amplitude, $\A(F_i\gets\l_1\gets I)$, and not on what happens in the other parts of the setup. 
 After $K\gg 1$ trials, 
 the experimenter counts the number of times, $dK(F_i,f)$, the system 
 is found in $|F_i\ra$ and the pointer's reading lies in an interval $df$ around $f$. 
 He can calculate a kind of \emph{average reading }
 \begin{equation}
\label{d1}
\overline{ f(F_i)}\equiv \int_0^K f dK(F_i,f)/K. 
\end{equation}
(It would be more natural to calculate the conditional average 
 by dividing $dK(F_i,f)$ by the number of successful post-selections in $|F_i\ra$
 $K(F_i)=\int dK(F_i,f)$. 
 This would yield Eqs.(\ref{a0}) and (\ref{0}) for the \emph{strong} and \emph{weak} regimes, respectively.
 However, the choice made in Eq.\ref{d1} is more convenient for the point we are trying to make.)

In the accurate \emph{strong} limit one has [cf.~\hyperref[appa]{Appendix~A}]
 \begin{equation}
\label{d2}
\overline{ f(F_i)} {\xrightarrow[ \Delta f \to 0 ] {}} |\A(F_i\gets \l_1\gets I)|^2,
\end{equation}
 clearly a local quantity which depends only on the amplitude of the route passing through the state $|\l_1\ra$, 
upon which $\B$ projects. In the inaccurate \emph{weak} limit the result
 \begin{align}
\label{d2a}
 \overline{ f(F_i)} {\xrightarrow[ \Delta f \to \infty ] {}} & \R\Big [\A(F_i\gets \l_1 \gets I)\times \n
& ~~~(\A^*(F_i\gets \l_1\gets I)+\A^*(F_i\gets \l_2\gets I))\Big ],
\end{align}
is non-local in the above sense, due to the presence of the second amplitude $\A^*(F_i\gets \l_2\gets I)$.
An attempt to probe the system locally while perturbing it only slightly, ends up probing 
all the pathways, leading to the same final condition. 
The orthodox may expect this from the Uncertainty Principle \cite{FeynL,FeynC}.
There is, however, still one possibility left. 

\subsection { Null weak value of a projector indicates absence of the system from the chosen location (?)}

 Finally, this seems to be a safe option (for more discussion see \cite{WVnull}), since $\la \B \ra_W$ in Eq.(\ref{b1}) can vanish only if the amplitude 
 $A(F_i\gets \l_1\gets I)$ is itself zero. With only one path, $F_i\gets \l_2\gets I$, leading to the final state, 
 there is no interference to destroy. The \emph{strong} and the \emph{weak} measurements apparently agree in that 
 the system never travels the route $F_i\gets \l_1\gets I$. 
 However, there is still a problem. To discuss it we will have to leave the double slit case, and increase 
 the number of routes, leading to the final state, to at least three. 

\section{A conjuring trick}
\label{sec:8}

Consider next a system in a three-dimensional Hilbert space, $N=3$, pre- and post-selected as before in states $|I\ra$ and $|F_i\ra$, respectively.
Three projectors, 
 \begin{gather}
\label{e0}
\B_1 =\hat \pi_1=|\l_1\ra\la \l_1|,\q \B_2 =\hat \pi_2=|\l_2\ra\la \l_2|, \n
\B_{1\cup2} =\hat \pi_{1\cup 2}=|\l_1\ra\la \l_1|+|\l_2\ra\la \l_2|
\end{gather}
monitor the presence of the system in the paths $F_i\gets \l_1\gets I$, $F_i\gets \l_2\gets I$, as well as in their union. 
The states $|I\ra$, $|\l_j\ra$ and $|F_i\ra$ are chosen to ensure that one has
 \begin{gather}
\label{e1}
\A(F_i\gets \l_1\gets I)= - \A(F_i\gets \l_2\gets I)\equiv A\n
\A(F_i\gets \l_3\gets I)\equiv A'. 
\end{gather}
The accurate \emph{strong} mean values of the three projectors, measured together, are [cf. Eq.(\ref{a0})]
 \begin{gather}
\label{e2}
\la \B_1\ra =\la \B_2\ra= \frac{|A|^2}{2|A|^2+|A'|^2}, \n
\la \B_{1\cup2}\ra =\la \B_1\ra+\la \B_2\ra= \frac{2|A|^2}{2|A|^2+|A'|^2}.
\end{gather}
This agrees with one's understanding of the concept of \emph{absence}. If a system (or at least a given property of that system) is absent from the union of two paths, $\la \B_{1\cup2}\ra=0$, it is because 
the system never takes either of them, $A=0$, and $\la \B_1\ra =\la \B_2\ra=0$. 

However, using Eq.(\ref{0}), for the \emph{weak} values one finds
 [cf. Eq.(\ref{a1})]
 \begin{gather}
\label{e3}
\la \B_1\ra_W =-\la \B_2\ra_W= \frac{A}{A'}, \n
\la \B_{1\cup2}\ra_W =\la \B_1\ra_W+\la \B_2\ra_W= 0.
\end{gather}
Unlike Eqs.(\ref{e2}), Eqs.(\ref{e3}) do not make conventional sense 
if the criterion (D) of the previous section is, indeed, valid. The system, \emph{not absent from the parts} (since $\la \B_1\ra_W, \la \B_2\ra_W \ne 0$)
is nevertheless \emph{absent from the whole} (since $\la \B_{1\cup2}\ra_W$=0). How can it be?
At this point the reader is expected to choose. 

One option is to see this as a \emph{paradox} peculiar to the bizarre quantum world. 
(In the same vein, by choosing in (\ref{e3}) $A=A'$, one can claim that \emph{a particle can be found with certainty in two different boxes} \cite{WV3b}, 
or that \emph{photons have discontinuous trajectories} \cite{WVph})

The other view is that there is no paradox, since the proposition (D) is simply wrong, and could, indeed, be expected to fail. 
One can only learn about the past scenario by destroying interference where various 
scenarios interfere. A weak pointer does not destroy it and, therefore, ceases to be a valid measuring device. 
Equations (\ref{e3}) only reflect the correct relations between the three amplitudes in Eq.(\ref{e1}), already known 
from the moment the choice (\ref{e1}) was made. There is no way around the Uncertainty Principle. 

Our main point is that the second view must be the right one.
To an orthodox the first option is an obvious fallacy. A proposition, based on a well defined concept, is found to contradict the evidence, the concept is amplified to encompass the evidence, so that the evidence can be seen as supporting the proposition. 
A meaningful explanation, or interpretation, is possible only if the standards against which a phenomenon is judged are 
maintained the same throughout the analysis. 
{What is wrong is that we do not ask what is right \cite{CHEST1}.}
 The need to constantly change the basic concepts in order to suit particular views, is yet another kind of trouble
 one finds inside Feynman's \emph{blind alley}.

\section{Summary}
\label{sec:9}

In a nutshell, the \emph{orthodox} view expressed in \cite{FeynL,FeynH,FeynC} can be summarized as follows.
The theorist's task is to combine known probability amplitudes as appropriate, so that the absolute square of the result would yeld the desired 
probability. 
(Needless to say, his/her other task involves constructing, in each case, a Hilbert space, a Hamiltonian, and the operators, which represent the
measured quantities.) 
This simple prescription, however, conceals a \emph{paradox} \cite{FeynC}, or a \emph{mystery} \cite{FeynL}.
The knowledge of the amplitudes alone cannot be used to determine a quantum system's past.
This is one way to state the Uncertainty Principle \cite{FeynL}.
Feynman recommended adjusting one's feelings about reality to reality, and strongly advised 
against rationalizing the quantum law using classical analogies.
An often cited Feynman's quote reads:
\begin{quote}
The `paradox' is only a conflict between reality and your feeling of what reality `ought to be.'
\cite[\S 18-3]{paradox}
\end{quote}

\noindent What may happen if this advice is ignored was illustrated by the three-slit case of Section~\ref{sec:8}. 
Accepting the premise that a vanishing amplitude signals the absence of the system from a path (or a \emph{box} \cite{WV3b}),
may lead to a \emph{paradoxical} notion that a quantum system may be \emph{present in the parts, yet absent from the whole}.
To an orthodox, there is no paradox, but rather a proof that the premise was wrong.

This example underlines the main difficulty in finding a physical meaning for \emph{weak values} in Eqs.(\ref{0}), 
given by combinations of the relative (i.e., normalised to a unit sum over the paths connecting the initial and final states) 
probability amplitudes. Like the amplitudes themselves, these combinations are always known to the theorist,
and the fact that their values can be deduced from the experimental data makes little difference to him. 
The question is whether \emph{weak values} can consistently describe the system's past in the presence of interference, 
something apparently forbidden by the Uncertainty Principle. 
 The orthodox believes that such a description is not possible, and the few propositions studied in Section~\ref{sec:7} appear to support this conclusion. 
If asked \emph{what is a `weak value'?} he can only refer to the probability amplitudes which quantum theory uses to describe the measured system, 
and cite the Uncertainty Principle as the main limitation on their possible use. 

Finally, the reader may ask whether all this nitpicking was really necessary. 
Surely the \emph{weak measurements} have practical uses, e.g., due to their amplifying effect \cite{WVtod}, 
so why deprive them of their allure? 
Why not allow for a bit of \emph{magic} where it helps to advertise the approach, or to gain a publication in a prestigious journal?
One answer is that, 
given the current interest in quantum technologies (some call it \emph{the second quantum revolution} \cite{Rev}), 
it is highly desirable to have full understanding of both possibilities and limitations of the basic theory which underpins 
the engineering developments.

\section*{Acknowledgment}

The author gratefully acknowledges financial support from Grant PID2021-126273NB-I00 funded by MICINN/AEI/10.13039/501100011033 and by ``ERDF A way of making Europe,'' and from the Basque Government Grant No. IT1470-22.

\section{Appendix A. Measurements, \emph{strong} and \emph{weak}}
\label{appa}

\subsection{Two-step measurement}

For our purpose it is sufficient to consider a system in a Hilbert space of a finite dimension $N\ge 2$.
The simplest measurement consists of preparing it in a state $|I\ra$ (step one) and measuring an operator $\B=\sum_{i=1}^N B_i|F_i\ra\la F_i|$, diagonal in an
orthonormal basis $\{|F_i\ra\}$, $i=1,2,...N$, after time $t_1$ (step two). According to \cite{FeynL}, there are $N$ scenarios, $F_i \gets I$, $N$ amplitudes, $\A(F_i\gets I)=\la F_i|\u(T)|I\ra$ [$\u(t)$ is the system's evolution operator] (see Fig.~\ref{fig_1}A). There are also $N$ probabilities, $P(F_i\gets I)=|\A(n\gets I)|^2$, since all the scenarios lead to distinguishable final conditions \cite{FeynL}. One can couple to the system a von Neumann pointer \cite{vN} with position $f$, prepared in a state $|G\ra$. 
The coupling Hamiltonian is 
$H_{int}=-i\partial_f\B \delta(t-t_1)$, 
and $G(f) \equiv \la f|G\ra$ is a real valued smooth function, peaked around $f=0$, 
with a characteristic width $\Delta f$ and zero mean, 
\begin{eqnarray}
\label{A-3}
 \int G(f)^2 df=1, \q \int fG(f)^2 df=0.
\end{eqnarray}
In each of the above scenarios the pointer's state is displaced by $B_i$. Thus, the probability to find a pointer's reading $f$ 
is simply 
\begin{eqnarray}
\label{A-2}
P(f\gets I,G) =\sum_{i=1}^N G^2(f-B_i) P(F_i\gets I). 
\end{eqnarray}
The \emph{mean} reading, 
\begin{align}
\label{A-1}
\la f(F_i)\ra &\equiv \frac{\int fP(F_i,f \gets I,G)df}{\int P(F_i,f \gets I,G)df}\n
&= \frac{\sum_{i=1}^N B_i P(F_i\gets I)}{\sum_{i=1}^N P(F_i\gets I)},
\end{align}
is, therefore, independent of the initial uncertainty in the pointer's position, $\Delta f$, which determines the accuracy of the measurement.
This, we note, is because all system's scenarios are \emph{a priori} endowed with probabilities, and there is no interference the pointer can destroy. This changes if more measurements are made.

\subsection{Three-step measurement}

Interesting interference effects first appear if the previously prepared system is to be measured twice, at $t_1$ in a basis $\{|\l_j\ra\}$, and then at $t_2$ in a basis $\{|F_i\ra\}$. There are $N^2$ scenarios, $F_i \gets \l_j \gets I$, and $N^2$ amplitudes, 
\begin{align}
\label{A0}
& \A(F_i \gets \l_j \gets I) =\la F_i|\u(t_2,t_1)|\l_j \ra\la \l_j|\u(t_1)|I\ra, \n
& \sum_{j=1}^N \A(F_i \gets \l_j \gets I)=\A(F_i \gets I).
\end{align}
If the system is left on its own, the scenarios interfere, and individual probabilities can be assigned only if this interference is destroyed \cite{FeynL}.
The measurement at $t=t_2$ accurately determines the system's final state $|F_i\ra$, but at 
 $t=t_1$ the measurement of an operator $\B=\sum_{j=1}^N B_j|\l_j \ra\la \l_j|$ can be fuzzy, $0 < \Delta f < \infty$. 
 It each scenario the pointer's state is shifted by $B_j$, and the transition amplitudes of the composite 
 \{system + pointer\} are particularly simple, 
\begin{eqnarray}
\label{A1}
\A(F_i,f \gets I,G) =\sum_{j=1}^N G(f-B_j)\A(F_i \gets \l_j \gets I). 
\end{eqnarray}
Now the joint probability of finding the system in $|F_i\ra$ at $t=t_2$, and having a pointer's reading $f$, is given by
\begin{eqnarray}\label{A1a}
P(F_i,f \gets I,G) = |\A(F_i,f \gets I,G)|^2, 
\end{eqnarray}
which reduces to $P(f\gets I,G)$ in Eq.(\ref{A-2}) if summed over all final states,
\begin{eqnarray}\label{A1b}
\sum_{i=1}^N P(F_i,f \gets I,G) = P(f\gets I,G), 
\end{eqnarray}
i.e., if the information about the measurement at $t_2$ is erased. 
The exact form of $P(F_i,f \gets I,G)$ now depends on how accurately $\B$ is measured. 
A highly accurate \emph{strong} measurement, $\Delta f \to 0$, destroys the interference, 
\begin{align}
& P(F_i,f \gets I,G) {\xrightarrow[ \Delta f \to 0 ] {}} 
\sum_{j=1}^N \delta (f-B_j)|\A(F_i \gets \l_j \gets I)|^2, \nonumber\\
& \label{A2}
\end{align}
and finding $f=B_j$ the experimenter knows that the system has traveled the route $F_i\gets \l_j\gets I$. 
The mean reading, conditional on the system arriving in $|F_i\ra$, $\la f(F_i)\ra \equiv \int f P(F_i,f \gets I,G)fd/\int P(F_i,f \gets I,G)df$ 
is, therefore, given by [cf. Eq.(\ref{a0})]
\begin{eqnarray}\label{A2a}
\la f(F_i)\ra {\xrightarrow[ \Delta f \to 0 ] {}}\frac {\sum_{j=1}^NB_j|\A(F_i \gets \l_j \gets I)|^2}{\sum_{j=1}^N|\A(F_i \gets \l_j \gets I)|^2}
=\la \B\ra.
\end{eqnarray}
In the opposite inaccurate \emph{weak} limit, $\Delta f \to \infty$, $P(F_i,f \gets I,G)$ is broad in $f$, interference is not destroyed, and 
an individual reading cannot identify the route taken by the system. 

 The role of the Uncertainty Principle \cite{FeynL} is best illustrated by treating the states $|\l_j\ra$ and $|F_i\ra$ as the \emph{slits}, and the 
\emph{points on the screen}, respectively. Now $P(F_i\gets I,G)\equiv \int P(F_i,f \gets I,G)df$ is the \emph{observed intensity}, 
which contains an \emph{inteference pattern}, provided $\Delta f \to \infty$
\begin{eqnarray}\label{A2b}
P(F_i\gets I,G){\xrightarrow[ \Delta f \to \infty ] {}}
 \left |\sum_{j=1}^N \A(F_i \gets \l_j \gets I)\right |^2\q \n
\ne\sum_{j=1}^N \left |\A(F_i \gets \l_j \gets I)\right |^2
{\xleftarrow[ \Delta f \to 0 ] {}}P(F_i\gets I,G).
\end{eqnarray}
In the \emph{weak} limit one can still evaluate $\la f(F_i)\ra$, and using $G(f-B_j) \approx G(f)-\partial_f G(f) B_j$
finds 
\begin{align}
& \la f(F_i)\ra {\xrightarrow[ \Delta f \to \infty ] {}}
\R\left[\frac{\sum_{j=1}^2 B_j\A(F_i \gets \l_j\gets I)}{\sum_{j=1}^2\A(F_i \gets \l_j\gets I)}\right ] 
\equiv \R[\la \B\ra_W]. \nonumber\\
& \label{A4}
\end{align}
Measuring the pointer's mean momentum, $\la \lm\ra$, allows one to determine its imaginary part (see, e.g., \cite{DSpl1})
\begin{equation}
\label{A5}
\la \lm(F_i) \ra 
{\xrightarrow[ \Delta f \to \infty ] {}}2\int \lm^2 G(\lm)^2 d\lm \times \Im[\la \B\ra_W].
\end{equation}
where $G(\lm)\equiv \la \lm|G\ra$ , and $\la f|\lm\ra =\exp(i\lm f)$.

After two experiments, each involving many trials, the experimenter determines the value of a complex quantity $\la \B\ra_W$. It is up to the theorist to explain what exactly has been learned about the observed system.

\section{Appendix B. Response of a system to a small perturbation}
\label{appb}

A two-level system is making a transition between states $|I\ra$ 
and $|F\ra$. The probability to detect it in $|F\ra$ after a time $t$ is
$P_0 =|\A_0|^2\equiv |\la F|\u(t)| I\ra|^2$.
A small perturbation $\hat V(t) =[V_1 |\l_1\ra\la \l_1|+V_2 |\l_2\ra\la \l_2|]\delta(t-t')$, 
introduced at $0<t'<t$, changes the transition amplitude to 
$\A \approx \A_0-i\delta \A$,
with $\delta \A=\sum_{j=1}^2 V_j\la F|\u(t,t')| \l_j\ra\la \l_j|\u(t')| I\ra$
The probability of detection then changes to $P_0 +\delta P $, 
where
$\delta P \approx 2\Im[\A^*_0\delta \A]$.
From the measured percentage change in the detection probability one can deduce the imaginary part of the 
\emph{weak value} of the perturbation $\la \hat V\ra_W$ [cf. Eq.(\ref{0})], 
\begin{eqnarray}\label{B0}
\frac{\delta P}{2P_0}\approx \Im \left[
 \frac{\sum_{j=1}^2 V_j\A(F_i \gets \l_j\gets I)}{\sum_{j=1}^2\A(F_i \gets \l_j\gets I)}\right ]
 = \Im\left [\la \hat V\ra_W\right ].
\end{eqnarray}

\section{Appendix C. Transition amplitudes for a two-level system in Fig.~\ref{fig_1}B}
\label{appc}

Without loss of generality, we choose the initial state of the spin to be polarized along 
the $z$-axis, $|I\ra =|z_+\ra$, the states at $t_1$ to be polarized along a direction 
$\vec n =(\sin\theta \cos\phi, \sin\theta \sin\phi, \cos\theta)$, $|\l_{1,2}\ra=|n_\pm\ra$, 
and the final states polarized up and down an axis $\vec n' =(\sin\theta' \cos\phi', \sin\theta' \sin\phi', \cos\theta')$, 
$|F_{1,2}\ra=|n'_\pm\ra$. System's evolution, if any, can be absorbed in the states $|\l_{1,2}\ra$ and 
$|F_{1,2}\ra$, so we also choose $\u(t)=1$. The four amplitudes are
\begin{eqnarray}\label{C0}
\A(F_1\gets \l_1\gets I)= \{\cos (\theta'/2)\cos (\theta/2)+\q \n
\exp[i(\phi-\phi']\sin (\theta'/2)\sin (\theta/2)\}\cos(\theta/2),\n
\A(F_1\gets \l_2\gets I)= \{\cos (\theta'/2)\sin (\theta/2)-\q \n
\exp[i(\phi-\phi')]\sin (\theta'/2)\cos (\theta/2)\}\sin(\theta/2),\n
\A(F_2\gets \l_1\gets I)=\{\sin(\theta'/2)\cos (\theta/2)-\q \n
\exp[i(\phi-\phi')]\cos (\theta'/2)\sin(\theta/2)\}\cos(\theta/2),\n
\A(F_2\gets \l_2\gets I)= \{\sin (\theta'/2)\sin (\theta/2)+\q \n
\exp[i(\phi-\phi')]\cos (\theta'/2)\cos (\theta/2)\}\sin (\theta/2).
\end{eqnarray}
Adding them up yields the two amplitudes in Fig.~\ref{fig_1}A, 
\begin{eqnarray}\label{C1}
\A(F_1\gets I)=\sum_{j=1,2}\A(F_1\gets \l_j\gets I)= \cos (\theta'/2)\n
\A(F_2\gets I)=\sum_{j=1,2}\A(F_2\gets \l_j\gets I)= \sin (\theta'/2).
\end{eqnarray}
For $\theta=\theta' =\pi$ and $\phi'=0$, the \emph{weak value} of the projector in Eq.(\ref{b0}) is
complex valued, with no restriction on the signs of its real and imaginary parts, 
\begin{eqnarray}\label{C2}
\la \B\ra_W=\la \hat \pi _1\ra_W=\cos\phi +i\sin \phi. 
\end{eqnarray}
For $\theta'=\pi-\epsilon$, $\epsilon \to 0$ and $\theta\ne \pi$, $|F_1\ra$ becomes a \emph{dark state}
(akin to a dark fringe in the interference pattern). 
The \emph{weak value} tends to 
\begin{eqnarray}\label{C3}
\la \B\ra_W=\la \hat \pi _1\ra_W= \to \exp[i(\phi-\phi')]\frac{\cos\theta}{\epsilon},\q 
\end{eqnarray}
and $|\la \B\ra_W|$ can exceed unity. For example,
putting $\phi=\phi'$, $\theta=0$, yields ($\hat \pi _2=|\l_2\ra\la \l_2|=1-\hat \pi _1$)
\begin{eqnarray}\label{C3}
\la \hat \pi _1\ra_W \to{1}/{\epsilon},\q
 \la \hat \pi _2\ra_W \to{1 -1}/{\epsilon}. 
\end{eqnarray}

\balance


\begin{thebibliography}{10}
\expandafter\ifx\csname url\endcsname\relax
  \def\url#1{\texttt{#1}}\fi
\expandafter\ifx\csname urlprefix\endcsname\relax\def\urlprefix{URL }\fi
\expandafter\ifx\csname href\endcsname\relax
  \def\href#1#2{#2} \def\path#1{#1}\fi

\bibitem{DSqua}
D.~Sokolovski. Are the weak measurements really measurements?. \emph{Quanta}
  2013; \textbf{2}:50--57.
\newblock \href {http://doi.org/10.12743/quanta.v2i1.15}
  {\path{doi:10.12743/quanta.v2i1.15}}.

\bibitem{WV1}
Y.~Aharonov, E.~Cohen, A.~Landau, A.~C. Elitzur. The case of the disappearing
  (and re-appearing) particle. \emph{Scientific Reports} 2017;
  \textbf{7}(1):531.
\newblock \href {http://doi.org/10.1038/s41598-017-00274-w}
  {\path{doi:10.1038/s41598-017-00274-w}}.

\bibitem{WV2}
Y.~Aharonov, E.~Cohen, S.~Popescu. A dynamical quantum {C}heshire {C}at effect
  and implications for counterfactual communication. \emph{Nature
  Communications} 2021; \textbf{12}(1):4770.
\newblock \href {http://doi.org/10.1038/s41467-021-24933-9}
  {\path{doi:10.1038/s41467-021-24933-9}}.

\bibitem{WVcrap}
C.~Ferrie, J.~Combes. How the result of a single coin toss can turn out to be
  $100$ heads. \emph{Physical Review Letters} 2014; \textbf{113}(12):120404.
\newblock \href {http://doi.org/10.1103/PhysRevLett.113.120404}
  {\path{doi:10.1103/PhysRevLett.113.120404}}.

\bibitem{WVcat}
Y.~Aharonov, S.~Popescu, D.~Rohrlich, P.~Skrzypczyk. Quantum {C}heshire {C}ats.
  \emph{New Journal of Physics} 2013; \textbf{15}(11):113015.
\newblock \href {http://doi.org/10.1088/1367-2630/15/11/113015}
  {\path{doi:10.1088/1367-2630/15/11/113015}}.

\bibitem{FeynL}
R.~P. Feynman, R.~B. Leighton, M.~L. Sands. Quantum behavior. in: The Feynman
  Lectures on Physics. Volume III. Quantum Mechanics. California Institute of
  Technology, Pasadena, California, 2013.
\newblock \url{https://www.feynmanlectures.caltech.edu/III_01.html}.

\bibitem{CHEST}
G.~K. Chesterton. Heretics. John Lane, London, 1905.
\newblock \url{https://archive.org/details/heretics01chesgoog}.

\bibitem{FeynP}
R.~P. Feynman. Epaulettes and the {P}ope. in: The Pleasure of Finding Things
  Out. Perseus Books, Cambridge, Massachusetts, 1999. pp. 7--8.

\bibitem{FeynC}
R.~P. Feynman. The Character of Physical Law. MIT Press, Cambridge,
  Massachusetts, 1985.

\bibitem{Hidd}
N.~D. Mermin. Hidden variables and the two theorems of {J}ohn {B}ell.
  \emph{Reviews of Modern Physics} 1993; \textbf{65}(3):803--815.
\newblock \href {http://doi.org/10.1103/RevModPhys.65.803}
  {\path{doi:10.1103/RevModPhys.65.803}}.

\bibitem{Cop}
J.~Faye. Copenhagen interpretation of quantum mechanics. in: E.~N. Zalta (Ed.),
  Stanford Encyclopedia of Philosophy. Stanford University, Stanford,
  California, 2019.
\newblock \url{https://plato.stanford.edu/entries/qm-copenhagen/}.

\bibitem{Bohm}
S.~Goldstein. Bohmian mechanics. in: E.~N. Zalta (Ed.), Stanford Encyclopedia
  of Philosophy. Stanford University, Stanford, California, 2021.
\newblock \url{https://plato.stanford.edu/entries/qm-bohm/}.

\bibitem{DSadp}
D.~Sokolovski, D.~Alonso~Ramirez, S.~Brouard~Martin. Speakable and unspeakable
  in quantum measurements. \emph{Annalen der Physik} 2023;
  \textbf{535}(10):2300261.
\newblock \href {http://doi.org/10.1002/andp.202300261}
  {\path{doi:10.1002/andp.202300261}}.

\bibitem{WVtod}
Y.~Aharonov, S.~Popescu, J.~Tollaksen. A time-symmetric formulation of quantum
  mechanics. \emph{Physics Today} 2010; \textbf{63}(11):27--32.
\newblock \href {http://doi.org/10.1063/1.3518209}
  {\path{doi:10.1063/1.3518209}}.

\bibitem{ABL}
Y.~Aharonov, P.~G. Bergmann, J.~L. Lebowitz. Time symmetry in the quantum
  process of measurement. \emph{Physical Review} 1964;
  \textbf{134}(6B):B1410--B1416.
\newblock \href {http://doi.org/10.1103/PhysRev.134.B1410}
  {\path{doi:10.1103/PhysRev.134.B1410}}.

\bibitem{Matz2}
A.~Matzkin. Weak values and quantum properties. \emph{Foundations of Physics}
  2019; \textbf{49}(3):298--316.
\newblock \href {http://doi.org/10.1007/s10701-019-00245-3}
  {\path{doi:10.1007/s10701-019-00245-3}}.

\bibitem{WVneg}
Y.~Aharonov, S.~Popescu, D.~Rohrlich, L.~Vaidman. Measurements, errors, and
  negative kinetic energy. \emph{Physical Review A} 1993;
  \textbf{48}(6):4084--4090.
\newblock \href {http://doi.org/10.1103/PhysRevA.48.4084}
  {\path{doi:10.1103/PhysRevA.48.4084}}.

\bibitem{DSx}
X.~G. de~la Cal, M.~Pons, D.~Sokolovski. Speed-up and slow-down of a quantum
  particle. \emph{Scientific Reports} 2022; \textbf{12}(1):3842.
\newblock \href {http://doi.org/10.1038/s41598-022-07599-1}
  {\path{doi:10.1038/s41598-022-07599-1}}.

\bibitem{WVh}
Y.~Aharonov, A.~Botero, S.~Popescu, B.~Reznik, J.~Tollaksen. Revisiting
  {H}ardy's paradox: counterfactual statements, real measurements, entanglement
  and weak values. \emph{Physics Letters A} 2002; \textbf{301}(3):130--138.
\newblock \href {http://doi.org/10.1016/S0375-9601(02)00986-6}
  {\path{doi:10.1016/S0375-9601(02)00986-6}}.

\bibitem{WVnull}
Q.~Duprey, A.~Matzkin. Null weak values and the past of a quantum particle.
  \emph{Physical Review A} 2017; \textbf{95}(3):032110.
\newblock \href {http://doi.org/10.1103/PhysRevA.95.032110}
  {\path{doi:10.1103/PhysRevA.95.032110}}.

\bibitem{WV3b}
T.~Ravon, L.~Vaidman. The three-box paradox revisited. \emph{Journal of Physics
  A: Mathematical and Theoretical} 2007; \textbf{40}(11):2873--2882.
\newblock \href {http://doi.org/10.1088/1751-8113/40/11/021}
  {\path{doi:10.1088/1751-8113/40/11/021}}.

\bibitem{WVph}
A.~Danan, D.~Farfurnik, S.~Bar-Ad, L.~Vaidman. Asking photons where they have
  been. \emph{Physical Review Letters} 2013; \textbf{111}(24):240402.
\newblock \href {http://doi.org/10.1103/PhysRevLett.111.240402}
  {\path{doi:10.1103/PhysRevLett.111.240402}}.

\bibitem{CHEST1}
G.~K. Chesterton. What's Wrong with the World. Cassell and Company, London,
  1910.
\newblock \url{https://archive.org/details/whatswrongwithwo03chesuoft}.

\bibitem{FeynH}
R.~P. Feynman, A.~R. Hibbs. Quantum Mechanics and Path Integrals. McGraw-Hill,
  New York, 1965.

\bibitem{paradox}
R.~P. Feynman, R.~B. Leighton, M.~L. Sands. Angular momentum. in: The Feynman
  Lectures on Physics. Volume III. Quantum Mechanics. California Institute of
  Technology, Pasadena, California, 2013.
\newblock \url{https://www.feynmanlectures.caltech.edu/III_18.html}.

\bibitem{Rev}
J.~P. Dowling, G.~J. Milburn. Quantum technology: the second quantum
  revolution. \emph{Philosophical Transactions of the Royal Society of London
  Series A: Mathematical, Physical and Engineering Sciences} 2003;
  \textbf{361}(1809):1655--1674.
\newblock \href {http://doi.org/10.1098/rsta.2003.1227}
  {\path{doi:10.1098/rsta.2003.1227}}.

\bibitem{vN}
J.~von Neumann. Mathematical Foundations of Quantum Mechanics. Princeton
  University Press, Princeton, 1955.
\newblock \href {http://doi.org/10.23943/princeton/9780691178561.001.0001}
  {\path{doi:10.23943/princeton/9780691178561.001.0001}}.

\bibitem{DSpl1}
D.~Sokolovski. Weak measurements measure probability amplitudes (and very
  little else). \emph{Physics Letters A} 2016; \textbf{380}(18-19):1593--1599.
\newblock \href {http://doi.org/10.1016/j.physleta.2016.02.051}
  {\path{doi:10.1016/j.physleta.2016.02.051}}.

\end{thebibliography}
\end{document}